%% file: llncs.tex
\documentclass{llncs}

\usepackage[english]{babel}
\usepackage[utf8]{inputenc}
\usepackage{amsmath,amssymb}
\usepackage{graphicx}
\usepackage{float}%figure environment
\usepackage[table,xcdraw]{xcolor}
\usepackage[disable]{todonotes}
\usepackage{booktabs}
\usepackage{wrapfig}
\usepackage{multirow}
\usepackage{tikz}
\usetikzlibrary{positioning,shapes,arrows}
\usepackage[ruled,vlined]{algorithm2e}
\usepackage{comment}
\usepackage{times}
\usepackage{rotating}
\newcommand*\rot{\rotatebox{90}}
\usepackage{longtable}

\newcommand{{\features}}{14\xspace}
\newcommand{{\systems}}{6\xspace}
\newcommand{{\classifiers}}{16\xspace}

\begin{document}

\title{Using Multi-Label Classification for Improved Question Answering}
%AN: multi-label => Multi-label as it is the most common way to write it
%\numberofauthors{5} 
\begin{comment}
\author{
\alignauthor
Ricardo Usbeck\\
       \affaddr{DICE Group, Paderborn University, Germany}\\
       \email{ricardo.usbeck@uni-paderborn.de}
\alignauthor
Michael Hoffmann\\
       \affaddr{DICE Group, Paderborn University, Germany}\\
       \email{hoffmann@informatik.uni-leipzig.de}
\alignauthor
Michael Röder\\
       \affaddr{DICE Group, Paderborn University, Germany}\\
       \email{michael.roeder@uni-paderborn.de}
\and  
\alignauthor
Jens Lehmann\\
       \affaddr{University of Bonn, Germany}\\
       \affaddr{Fraunhofer IAIS, Germany}\\ \email{jens.lehmann@\\cs.uni-bonn.de}
\alignauthor
\mbox{Axel-Cyrille Ngonga Ngomo}\\
       \affaddr{DICE Group, Paderborn University, Germany}\\
       \email{ngonga@uni-paderborn.de}
}
\lstset{  
    escapeinside={(*}{*)}
}
\end{comment}

\author{
Ricardo Usbeck\inst{2} \and
Michael Hoffmann\inst{1} \and
Michael Röder\inst{2} \and 
Jens Lehmann\inst{3} \and
Axel-Cyrille Ngonga Ngomo\inst{2} 
}

\institute{
Leipzig University, Germany 
\and
University of Paderborn, Germany
\and 
University of Bonn, Germany and Fraunhofer IAIS, Germany 
\newline email: \{firstname.lastname\}@uni-paderborn.de}

\maketitle
\begin{abstract}
A plethora of diverse approaches for question answering over RDF data have been developed in recent years. While the accuracy of these systems has increased significantly over time, most systems still focus on particular types of questions or particular challenges in question answering.
What is a curse for single systems is a blessing for the combination of these systems. We show in this paper how machine learning techniques can be applied to create a more accurate question answering metasystem by reusing existing systems. 
In particular, we develop a multi-label classification-based metasystem for question answering over \systems existing systems using an innovative set of \features question features.
The metasystem outperforms the best single system by 14\% F-measure on the recent QALD-6 benchmark.
Furthermore, we analyzed the influence and correlation of the underlying features on the metasystem quality.
\end{abstract}

\todo[inline]{
- the related work section does not discuss meta classifiers for related fields with the usual techniques, such as bagging, boosting, as well as multi-label classification

- paper is unclear in important aspects, such as how the meta features are computed (some are explained later on in the implementation details)
  - for example, what does it mean "iff S\_i achieves an F1-score greater than 0 on q"
  - the construction of ranking+thr (RT) is unclear to me

- it is not clear if the results that are achieved are due to overfitting (unlike the author's statement that "the leave-one-out setup should converge to the second setup")

- I expected the authors to train and test on different sets, e.g. train on QALD 1-5 and test on QALD 6

- is the improvement compared to the best single system statistically significant? it seems that this single system contributes to most of the meta classifier results. is this true?

- In addition, IBM's Watson is taking a similar approach if I recall correctly, using several lite qa systems and combining evidence, instead of selecting the best one. This should be compared and discussed. The results themselves are not clear, the dataset seems small and statistical significance is missing.

Novelty is limited
* Related areas such as question classification in QA systems or query classification for meta-search engines are totally ignored in related works

Lack of clarity in the experimental section
* The data set is not described in enough details
* It is unclear whether training and test data are kept separate; especially sections 5.3 (F1-Score Full), 5.4 and 5.5

Poor use of space
* Is Table 4 really necessary?
* 25\% of citations are self-citations

* The idea classification in a meta-system is not novel; however, to the best of my knowledge, this is a first application of the concept to QALD. We encourage the authors to discuss applications of the concept in other areas such as search (see doi:10.1016/S1389-1286(00)00059-1) in the related work section. 

* A related concept is question classification in question answering systems (see doi:10.3115/1072228.1072378 for instance), and this line of research is also ignored by the authors. The authors focus on selecting a QALD engine, while question classification in QA focuses on selecting the best components - it's a related concept and should be mentioned.

*  The related work section focuses on the egineering aspect (how to query systems programmatically), rather than the research aspect (the methods that make these systems different). 

* This reviewer is unsure of the validity of the reported experiments in Section 5, esp. in 5.4 and 5.5. 

* The authors state: "performed tests to choose the best performing classifier on our data". Which data? Is training data kept separate from test data?

* The authors do not give enough details about the data set. When they mention 100 test questions, it is unclear whether these test questions are used for testing and training. Looking at the QALD-6 data, there exists a larger set of training questions. However, it is unclear whether the authors used the training data.

* How many classification examples does QALD-6 correspond to for training and testing? Is this enough data to compare 16 classifiers, some of them related by design?

* Table 4 is long and not very informative. What should the take-away be for the reader? 

* It would have been more constructive to perform an in-depth analysis of the successes and failures, i.e., expand on the last sentence of section 5.5

* There are a few unusual phrasing Section 2, page 2:  a domain- as well as a domain-independent ontology

* QANUS is a not disclosed QA framework

-Evaluation may lead to wrong conclusions

-Needs better motivation for the proposed solution

* The motivation for multi-label classification is that multiple labels can be associated with the same instance, that is, multiple systems can given the correct answer for the same question. However, it seems that the classifiers are always returning only only system. Maybe a way to improve the results would be to combine multiple systems, and not to select one of them. That said, I think the paper would be strengthened if the author provide a deeper discussion about the solution proposed (ie., delegation vs. combination).

* I could not understanding what is "question word". It is not described in Section 3.1. However, it seems to be one of the most important features.

* The base systems are not discussed properly. Apparently, some of them were not published as an academic paper. Readers may be not familiar with these systems, and this makes it hard to understand the weaknesses and strengths of the QA systems. The paper lacks a proper discussion about this issue.

* My main issue with this paper is the evaluation. The dataset is seriously limited in size. The results may change drastically on other datasets. There is no statistical test to ensure the superiority of the systems. The choice of "Pst" seems strange to me, since it is probably the most prone to overfitting among the available choices (that is probably the reason why it was so good in "Full" and not so good in "LOO").

* 1. Using machine learning to combine multiple systems is not novel
* 2. Lack technical depth
* 3. Experiments are weak, using only 100 questions for both training and test. 
}

\todo[inline]{There are two new systems to GERBIL QA}
\section{Introduction}
Recent research on question answering (QA) over Linked Data and RDF has shown significant improvements of quality and efficiency in answering even complex questions \cite{qasurvey}. As a result of this research, a multitude of QA systems (see a.o. \cite{YodaQA,hawk,Xser,treo,SINA_WebSemantic,PowerAqua,tbsl}) have been proposed to tackle questions from different domains and of varying complexities. These systems rely on diverse approaches ranging from the transformation of questions into triple patterns \cite{PowerAqua} to hybrid question answering over both RDF and text \cite{hawk}. This has led to a tool landscape with approaches able to deal with particular aspects of questions well (e.g., \cite{tbsl} can deal with simple conjunctive queries) while being unable to deal with other aspects (e.g., \cite{PowerAqua} has difficulties dealing with superlative queries). 
In addition to monitoring the development of a large number of question answering approaches, we have also witnessed the creation of a large number of benchmarks and challenges. The latter have provided the possibility to analyse the strengths and weaknesses of many QA systems objectively (see, e.g., QALD \cite{qald6,qald5,qald4}).

The availability of both diverse approaches (e.g., approaches with different strengths and weaknesses) and of benchmarks (that allow evaluating these strengths and weaknesses) now suggests the possibility of creating ``metasystems'' for answering questions over Linked Data and RDF. 
Such a metasystem (1) integrates several QA systems. Given a question, it is then able to (2) select the most appropriate QA system to answer the said question from a set of questions. While the selection of the most appropriate QA system seem tedious, the main hypothesis of this work is that this selection can be carried out automatically using machine learning (ML) techniques.

In this work, we formulate the problem of the training of a metasystem for QA as a multi-label classification problem. %, labeling each question with the QA systems that are able to answer it.
%TODO AN: Read paragraph below
Here, we are interested in the choice of the best fitting classifier and the choice of machine learning features which are most descriptive.
In this paper, we present a multi-label classification-based metasystem for question answering over \systems existing systems using a novel set of \features question features.

Our contributions are as follows: 
%\begin{itemize}
%\item 
(1) We develop a set of \emph{\features novel features} for natural-language questions that is capable of characterizing the weak points of existing QA systems. 
%\item 
(2) We analyze \emph{\systems current QA systems} with respect to their performance and features to deduce future research directions and gain insights into the systems' performances. 
%\item 
(3) We analyze \emph{\classifiers classifiers} to find the best performing multi-label classification system for the task at hand. 
%\item 
(4) We implement and present a machine learning approach for combining these QA systems with \classifiers classifiers. This \emph{metasystem outperforms the state of the art} in the QALD-6 benchmark~\cite{qald6}. 
%\item 
(5) We optimize the set of features used for training the metasystem to conclude with a minimal set of meaningful question \emph{features boosting the quality of the metasystem by 4\%}.
%\end{itemize}

More information about the approach, source code and underlying data can be found in our project repository \url{https://github.com/AKSW/NLIWOD/tree/master/qa.ml}.

%https://slack-files.com/files-pri-safe/T02RQPZG8-F12S1SUBD/qald-6_results.pdf?c=1476107056-12f988e2acb68acdb03e75a38fa1fbbdbc36e4bf

\section{Related Work}

With the growing number of published QA systems, a search for an universal framework for reusing components began.
One of the earliest works is openQA~\cite{openQA} which is a modular open-source framework for implementing QA systems. 
openQA's main work-flow consists of four stages ({interpretation}, {retrieval}, {synthesis} and {rendering}) as well as adjacent modules ({context} and {service}) written as rigid Java interfaces.
The authors claim that {openQA} enables a conciliation of different architectures and methods.
QALL-ME~\cite{qallme} is another open source approach using an architecture skeleton for multilingual QA, a domain- as well as a domain-independent ontology.
The underlying SOA architecture features several web services which are combined into one QA system in a predetermined way.
Another system is the open source OAQA~\cite{oaqa}.
This system aims to advance the engineering of QA systems by following architectural commitments to components for a better interchangeability.
Using these shared interchangeable components OAQA is able to search the most efficient combination of modules for a task at hand.

QANUS~\cite{qanus} is a not disclosed QA framework for the rapid development of novel QA systems as well as a baseline system for benchmarking.
It was designed to have interchangeable components in a pre-seeded pipeline and comes with a set of common modules such as named entity recognition and part-of-speech tagging.
Both et al.~\cite{both2014service} described a first semantic approach towards coupling components together via RDF to tailor search pipelines using semantic, geospatial and full text search modules.
Here, modules add semantic information to a query until the search can be solved. 
QANARY~\cite{singhqanary} is the first real implementation of a semantic approach towards the generation of QA systems from components. 
Using the provided QA ontology from QANARY, modules can be exchanged, e.g., various versions of NER tools, to benchmark various pipelines and choose the most performant one.

\todo[inline]{Verweis auf arxiv von self-wiring paper}

However, none of the frameworks is able to combine various QA systems.
To the best of our knowledge, we present the first system able to combine several QA systems based on question features which outperforms every single system performance.

\section{Approach}
\label{sec:approach}

In general, QA systems perform well on certain question domains like geography, physics, encyclopedic knowledge or particular knowledge source combinations~\cite{qasurvey}.
Our goal is to provide a metasystem which is able to pick the most capable, specialised QA system for a particular question. 
We formalize our problem as follows:
Let $q$ be a question (i.e., an instance in ML terminology), and $S_1, \dots, S_n$ be an enumeration of the QA system that underlies our metasystem. We  label $q$ by the vector $L(q) = (l_1, \dots, l_n)$, with $l_i = 1$ iff $S_i$ achieves an F1-score greater than 0 on $q$. Otherwise, we set $l_i = 0$.  
Given an unseen $q'$, our goal is to choose a QA system $S_j$ with the highest F1-score.
The problem at hand clearly translates to a  multi-label classification problem~\cite{tsoumakas2011random}.

The goal of multi-label classification is as follows: Given an unseen instance $q'$, assign one or more possible labels $L(q')$ to $q'$, where each label can have multiple classes.
In our case, we use a Boolean set of classes to indicate whether a system is able to answer a certain question or not.
Approaches to tackle multi-label classification in this form can be divided into two categories. 
The first one is to transform the multi-label problem into one or several single-label problems, i.e., training a separate classifier for each subproblem~\cite{tsoumakas2011random}. 
Depending on the algorithm, the next step could be a voting scheme or other methods to combine the separate classifications~\cite{tsoumakas2011random}. 
In our case, most classifiers fall into this category and are explained in detail in Section~\ref{sec:classifiers}.
The second category contains algorithm adaption methods, where one adapts existing machine learning algorithms to handle multi-label data directly. 
Examples of this method include Adaboost.MH/MR~\cite{Schapire:2000:BBS:347709.347732} or ML-kNN~\cite{zhang2007ml}. 
For an exhaustive overview of the techniques of multi-label classification, we refer to the survey of Tsoumakas et al.~\cite{tsoumakas2011random}.

\begin{figure}[htb]
\centering
\includegraphics[width=\linewidth]{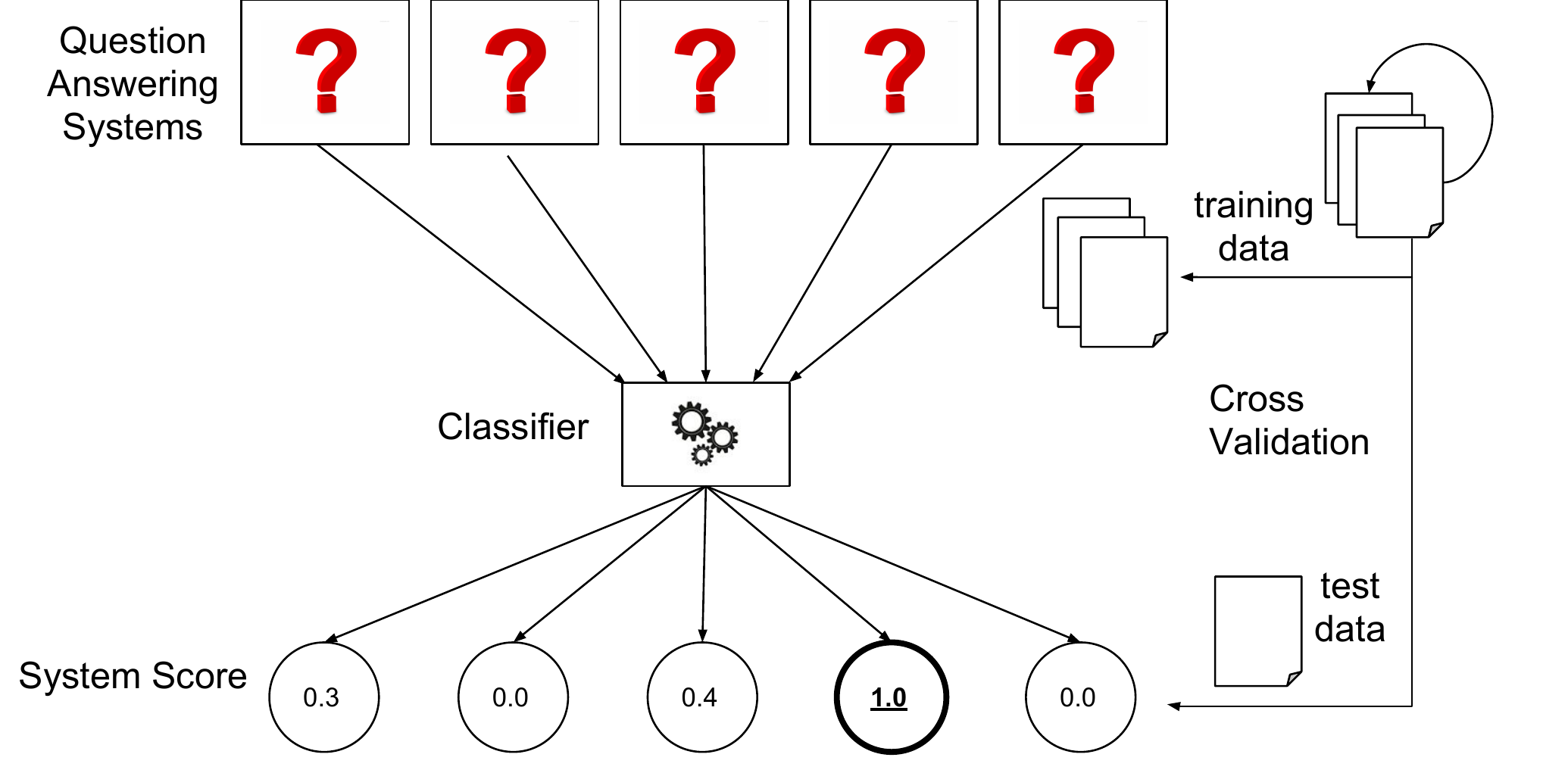}
\caption{Overview of the metasystem combining several QA systems. The training and test data are the feature vectors extracted from QALD-6 questions.}
\label{fig:overview_single_classification}
\end{figure}

Multi-label classification can be tackled using classical ML techniques~\cite{MEKA} provided that corresponding features are designed.  
We address this challenge in Section~\ref{sec:features}.
Using these features, we train a classifier (i.e., a metasystem) to select the system(s) that is/are most likely to be able to answer $q'$. 
We interpret the output of this classifier as a ranking among systems and query the system with the highest rank. 
%We will test all classifiers in the MEKA system, using 10-fold cross-validation to select the most likely to succeed.
Our overall approach is depicted in Figure~\ref{fig:overview_single_classification}.

%\subsection{multi-label Classification}

\subsection{Features}
\label{sec:features}
The \features features developed herein are based on recent surveys~\cite{qasurvey} as well as on an analysis of the results of previous question answering challenges~\cite{tbsl,hawk}.
These features can be summarized into eight groups. 
We explain each feature using the following running example: "\textit{Which New York Nicks players from outside the USA are born after Robin Lopez?}".
\begin{enumerate}
\item \textbf{Question Type:} This feature has four dimension, i.e., \texttt{List}, \texttt{Boolean}, \texttt{Resource} and \texttt{Number} and determines the type of the answer set. For our running example, the feature would take the value \texttt{List}.
\item \textbf{Query Resource Type:} With its seven dimensions, the feature categorizes each entry in the answer set to one of the following items: \texttt{Misc.}, \texttt{Date}, \texttt{Boolean}, \texttt{Number}, \texttt{Person}, \texttt{Organization}, and \texttt{Location}. This feature would be set to \texttt{Person} for our running example.
\item \textbf{Wh-type:} Although simple, this feature is highly effective in determining a spectrum of capabilities of a QA system, e.g., whether the said system is able to construct SPARQL ASK queries.\footnote{\url{http://www.w3.org/TR/2013/REC-sparql11-query-20130321/\#ask}} Using the first two tokens of an input question, this feature's dimensions are \texttt{Who}, \texttt{Which}, \texttt{How}, \texttt{In which}, \texttt{What}, \texttt{When}, \texttt{Where} as well as \texttt{Ask}. 
Note that the \texttt{Ask} dimension summarizes different questions that demand the generation of SPARQL ASK queries~\cite{HAWK_NLIWOD_2015} as well as questions starting with "\textit{Give me}" or "\textit{Show me}".
Our running example would be assigned the value \texttt{Which} for this dimension.
\item \textbf{\#Token:}	The number of tokens is calculated based on already identified entities and noun phrases and ignores punctuation. 
For example, our tokenized running example would be \textit{[Which] [New York Nicks] [players] [from] [outside] [the] [USA] [are] [born] [after] [Robin Lopez]} and would result in a numerical value of 11. 
%\item \textbf{Limit (includes order by and offset):} Certain questions ask for only an example out of a result list or a superlative. Thus, this feature accounts for resulting SPARQL queries which contain a LIMIT or an OFFSET. For our running example this boolean feature is false.
\item \textbf{Comparative:} This feature describes whether a question uses a comparative adjective, e.g., higher, or comparative words such as than, after, before.
For our running example, this boolean feature is \texttt{true}.
\item \textbf{Superlative:}	Like the Comparative feature, the Superlative feature indicates the use of a superlative, like highest or best, and is \texttt{false} for the example question.
%\item \textbf{Person, Location, Organization, Misc:} 
\item \textbf{Entity types}:
This group of features includes seven boolean features: \texttt{Person}, \texttt{Money}, \texttt{Location}, \texttt{Percent},\\ \texttt{Organization}, \texttt{Date} and \texttt{Misc.}.
Each feature describes whether an entity of this type exists within the question of this particular type. 
Our running example question, would have \texttt{Organization} (New York Nicks), \texttt{Place} (USA) and \texttt{Person} (Robin Lopez) set to \texttt{true} while the remaining features would be set to \texttt{false}. 
\end{enumerate}

While these features are clearly handcrafted, 
we show their ability to effectively determine the question answering systems according to their capabilities as well as to accurately choose the correct system to answer a certain question in Section~\ref{sec:evaluation}. 
Note that our metasystem is flexible so that each feature can be extended, as can the number of features themselves, to adapt to new QA benchmarks or systems.

\subsection{Classifiers}\label{sec:classifiers}
To compute a metasystem model we evaluated \classifiers multi-label classifiers from the MEKA framework~\cite{MEKA}.
In the following, we give a short overview of the classifiers we used: 
\begin{itemize}
    \item \textbf{Label Combination (LC)}: This method treats each possible combination of labels as a class and uses a multi-class classifier for classification.
    \item \textbf{Ranking + Threshold (RT)}: Each example is copied once for each label and is assigned one label. On this augmented data, a multi-class classifier is trained. To make predictions a sample is mapped to a ranking of possible labels and gets assigned all labels above a threshold.
    \item \textbf{Classifier Chains (CC)}: Read et al.~\cite{read2011classifier} introduced this classification method, which uses binary classifiers for each label ordered in a chain, such that the classifier for the i-th label is conditioned on the previous $i-1$ classification results. Since the performance depends on the order of the labels, there are extensions~\cite{read2014efficient,cheng2010bayes} to find the best suited chain. We include the \textbf{PMCC, MCC, CC, CT, BR} and \textbf{BRq} classifiers in this section, since they belong to the same family of classifiers.
    \item \textbf{Random disjoint k-labelsets (RAkELd)}: In 2011, Tsou\-makas et al.~\cite{tsoumakas2011random} introduced RAkELd which randomly partitions the labelset $L$ into $\lfloor |L|/k \rfloor$ disjoint labelsets with at most k labels. For each disjoint labelset a classifier is trained, using the Label Combination (LC) method. To classify a new instance, the results of all classifiers are gathered. \textbf{RAkELo} is an extension that also incorporates overlapping k-labelsets. \textbf{HASEL} partitions according to the hierarchy defined by the dataset.
    \item \textbf{Conditional Dependency Networks (CDN)}: Guo et al~\cite{guo2011multi} present CDN. This classifier models the probabilities for each label as a densely connected conditional dependency network of binary classifiers. The network is trained using logistic regression. The prediction for an instance is obtained by using Gibbs Sampling on this network to obtain the approximate joint distribution and using MAP inference on this approximation. Another member of this family is the  \textbf{CDT} classifier.
    \item \textbf{Pruned Sets (PS)}: This transformation method was introduced by Read et al.~\cite{read2008pruned} as well and transforms the multi-label classification problem into a multiclass classification problem, just like LC but also prunes infrequent label combinations to combat over fitting. \textbf{PSt} introduces a threshold into the classification.
\end{itemize}

\section{Implementation details}
All implementation details (including ML, feature extraction and evaluation) can be found in our open-source repository. To calculate each feature, an in-depth analysis of the input question using Part-of-Speech tags, dependency parse trees, string matching and entity recognition respectively disambiguation is required.
We rely on the Stanford CoreNLP library~\cite{stanfordcorenlp} in our current implementation. The classification algorithms we rely on are implemented in MEKA \cite{MEKA}.

Given that our evaluation is to be carried out on QALD-6, we introduce the participating systems of the QALD-6/Task-1 challenge~\cite{qald6}.
Table~\ref{tab:qaldoverview} shows the involved systems and possible reasons for exclusion from our metasystem.\footnote{
Most systems do not have a webservice to work with.
For a list of QA systems with available web services, please go to our project repository at \url{https://github.com/AKSW/NLIWOD/blob/master/qa.systems/README.md}}. 
We successfully contacted all authors and asked for permission to use their challenge entries to test our approach (systems below midrule in Table~\ref{tab:qaldoverview}). 

\begin{table}[htb!]
\centering
    \begin{tabular}{@{}lccc@{}}
    \toprule
    \textbf{Engine}       & \multicolumn{1}{l}{\textbf{Reference}} & \multicolumn{1}{l}{\textbf{Webservice}?} & \textbf{Exclusion} \\                 \\ \midrule
    CANaLI      & \cite{mazzeo2016canali}       &  yes   & M     \\
    PersianQA   & -                             &  no      & L     \\
    \midrule
    UTQA (English)& \cite{utqa}                   &  no      &                     \\
    KWGAnswer   & -                             & no       &                      \\
    NbFramework & -                             &  no      &                      \\
    SemGraphQA  & \cite{beaumont2015semgraphqa} &   no     &                      \\
    UIQA WME    & -                             &   no     &  (M)                    \\
    \bottomrule
    \end{tabular}
    \caption{Systems that participated in the 6th QALD challenges. Note, having a publication is optional with QALD. \textbf{Exclusion} indicates the reason for exclusion of a system from our dataset. L means the system is not available for English, M indicates that human interaction is needed. }
    \label{tab:qaldoverview}
\end{table}

Four out of seven systems (i.e., KWGAnswer, NbFramework, PersianQA and UIQA) have no attached publication at the time of writing. Thus, we are unable to describe their inner mechanisms.  
Note, that the UIQA system participated in QALD-6 as an automatic system as well as a human-supported system. 
We use UIQA without manual entries (WME) for our evaluation.

SemGraphQA~\cite{beaumont2015semgraphqa} is an unsupervised and graph-based approach which also limits itself to questions requiring only DBpedia~\cite{jl_2014/swj_dbpedia} types.
First, the approach tries to match RDF resources to parts of the natural language question and builds a syntactic parse tree from dependency parsing.
Second, the resulting structure is transformed into various possible semantic interpretations, i.e, resolving ambiguities indirectly. 

UTQA~\cite{utqa} is a crosslingual QA system based on a language-specific chunker for porting, a maximum entropy model and an answer type prediction. 
Found ground entities, the predicted answer type as well as a semantic similarity are then used to find matching neighbouring entities.

For further information about QA systems and the state of the art, please refer to~\cite{qasurvey,DBLP:journals/semweb/LopezUSM11,Kolomiyets:2011}. 

\section{Evaluation}
\label{sec:evaluation}

The purpose of this evaluation was four-fold.
First, we aimed to analyze the correlation between certain features and the QALD-6 submission data to point out current weak points as well as future research directions for each QA system.
Second, we analyzed the set of available multi-label classifiers and performed tests to choose the best performing classifier on our data.
Third, we studied the performance of our novel metasystem for question answering.
Finally, we analyzed the features required to optimize the metasystem as well as their influence.

\subsection{Dataset}
The evaluation of this approach is based on the 6th edition of the Question Answering over Linked Data challenge (QALD-6)~\cite{qald6}.
The dataset contains 100 test questions and the answers for the respective systems on task 1, multilingual question answering. 

\subsection{Feature Association with System Performance}

First, to assert the descriptiveness of our features, we calculated Cramers' V-coefficient for each feature and a system's ability to answer a question.
To this end, the ability to answer was divided into the two classes "can answer" (F1-score > 0) or "cannot answer" (F1-score = 0).
Cramers' V is based on the Chi-squared statistics and is defined as follows:
\begin{equation}
    V = \sqrt[2]{\frac{\mathcal{X}^2}{n(k-1)}},
\end{equation}
with $k = \min(I, J)$, I and J being the number of rows and columns of the contingency matrix of our experiment.
To define $\mathcal{X}^2$, fix some feature and let $\pi_{ij}$ be the observed count of event $(a(i), b(j))$, with $a(1) :=$  "can answer" and $a(2) :=$ "cannot answer", and $b(j)$ the j-th state of the feature.
Based on this contingency matrix, let $n$ be the number of observations, $\pi_{*i} := \sum_{j=1}^{J} \pi_{ij}$ and $\pi_{j*} := \sum_{i=1}^{I} \pi_{ij}$, then one defines
\begin{equation}
    \mathcal{X}^2 := n\sum_{i=1}^{I}\sum_{j=1}^{J}\frac{(\pi_{ij} - \pi_{*i}\pi_{j*}/n)^2}{\pi_{*i}\pi_{j*}}.
\end{equation}
Cramers' V estimates the association of the features based on the observed contingency matrix, $V=0$ implying statistical independence and $V=1$ implying that both features are linearly dependent. 
\begin{figure}[htb!]
    \centering
    \includegraphics[width=\linewidth]{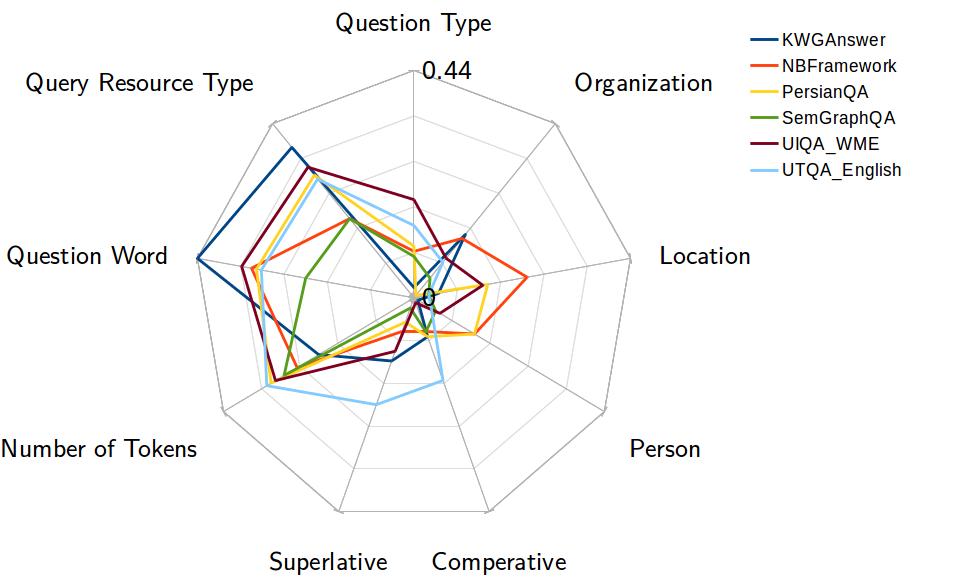}
    \caption{Cramers V-coefficient of features and the system performances.}
    \label{fig:spiderdiagram}
\end{figure}
Figure~\ref{fig:spiderdiagram} shows that across all QA systems the features \texttt{Query Resource Type}, \texttt{Question Word} and \texttt{Number of Tokens} demonstrate the closest association with a system's ability to answer. 
Furthermore, there seems to be a large association between the performance of NBFramework and \texttt{Location} (see for example questions 12, 20, 23 and 44 in Table 2). The same effect can be observed with UTQA (English) and \texttt{Superlative}, see questions 17, 18, 77. 
We investigate this effect further in Section~\ref{sec:featureinfluence}

%\todo[inline]{Figure \ref{fig:spiderdiagram} needs a larger labeling of its axis. The small $0.44$ is nearly not readable and looks like we are trying to hide it.}

\subsection{Choosing a Classifier}
\todo[inline]{leave-one-out statt cross validation}
Second, we determined the best classifier on all features.
To this end, we performed a 10-fold cross-validation for a set of classifiers $\text{C}_1, \dots \text{C}_m$, recording the macro F1-score on each fold. 
To be precise, we calculated the numbers:
\begin{equation}
    \text{Score}_i(\text{C}_j) = \frac{1}{|\text{Test-Fold}_i|}\sum_{q \in \text{Test-Fold}_i}\text{C}_j(q), 
\end{equation}
where we defined
\begin{eqnarray}
    \text{C}_j(q) &=& \text{F1-Score}(\text{S}_{\phi_j}(q)), \quad \text{with} \\
    \phi_j &=& \text{arg max} \{\text{Rank}_j(\text{S}_k) | k = 1, \dots, n \}. 
\end{eqnarray}

\begin{figure}[htb!]
    \centering
    \includegraphics[width=\linewidth]{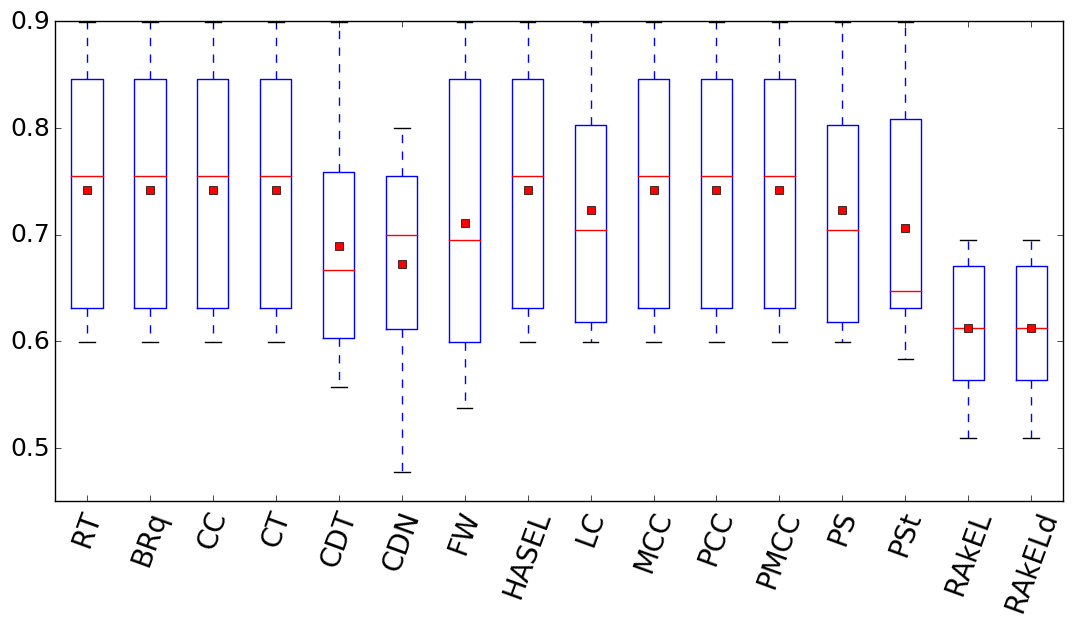}
    \caption{Boxplot of Cross-Validation results.}
    \label{fig:classifierperformance}
\end{figure}

Here, $\text{Rank}_j(\text{S}_k)$ refers to the rank, that the j-th classifier assigns to the k-th system and $S_i(q)$ is the set of answers provided by system i on question q.
We used the F1-scores according to the reported QALD-6 data for each system. 
%As a reminder, the F1-Score is defined as
%\begin{eqnarray}
%    \text{F1-score} &=& 2\frac{\text{Precision} \times \text{Recall}}{\text{Precision} + \text{Recall}},
%\end{eqnarray}
%with
%\begin{eqnarray}
%    \text{Precision}&=& \frac{\text{TP}}{\text{TP} + \text{FP}} \quad \text{and} \\
%    \text{Recall}   &=& \frac{\text{TP}}{\text{TP} + \text{FN}},
%\end{eqnarray}
%with respect to a set of provided answers. 
Figure~\ref{fig:classifierperformance} shows the results of our experiment in a boxplot.

The classifiers of the Classifier Chains (CC) family achieved the best performance on this task. 
However, we reached higher scores for a particular classifier if we used only a subset of features, see Section~\ref{sec:featureinfluence}.
%Here, we detected a lack of training data which cannot be circumvented due to missing queryable QA systems.

Furthermore, we tested the performance of all above classifiers, using all the features in two setups.
%In this experiment, we tested and trained on all questions, the results are displayed in Table~\ref{tab:classifier}.
First,  we tested and trained the classifiers on 99 questions and used the remaining single question for testing.
We repeated this leave-one-out procedure for all questions and calculated the average F1-score.
Second, we tested the performance of all above classifiers using all questions as training set.
The results are displayed in Table~\ref{tab:classifier}.

As can be seen, there is a huge difference between the two setups.
This shows the sparsity and vast diversity of the dataset that is caused by the low number of questions available.
With a growing number of questions, the results of the leave-one-out setup should converge to the second setup (i.e., F1-Score Full).
Surprisingly, the PSt classifier performed best on this task in the second setup and thus we chose it in our metasystem as it outperforms the best other classifier by 0.04 points F-measure. 

%We are aware that this choice could be induced by overfitting but we seek to leverage a feature selection to account for the sparsity of the training data.

\begin{table}[!htb]
    \centering
    \begin{tabular}{lcc}
    \toprule
    \textbf{Classifier} & \textbf{F1-Score leave-one-out} & \textbf{F1-Score Full} \\ 
    \midrule
    RT & 0.68 & 0.68\\
    BRq & 0.68 & 0.68\\
    CC & 0.68 & 0.68\\
    CT & 0.68 & 0.68\\
    CDT & 0.60 & 0.69\\
    CDN & 0.60 & 0.72\\
    FW & 0.68 & 0.68\\
    HASEL & 0.68 & 0.68\\
    LC & 0.66 & 0.71\\
    MCC & \textbf{0.69} & 0.68\\
    PCC & \textbf{0.69} & 0.68\\
    PMCC & \textbf{0.69} & 0.68\\
    PS & 0.66 & 0.71\\
    PSt & 0.63 & \textbf{0.76}\\
    RAkEL & 0.55 & 0.72\\
    RAkELd & 0.65 & 0.70\\
    \bottomrule
    \end{tabular}
    \caption{F1-Score leave-one-out: Classifier performance computed using Leave-One-Out methodology. F1-Score Full: Classifier performance tested and trained on all questions.}
    \label{tab:classifier}
\end{table}

\subsection{Feature Influence on Performance}
\label{sec:featureinfluence}
To probe the influence of the different features on the performance of the metasystem, we trained on all questions and the PSt classifier, using different combinations of features. 
We avoided using cross-validation due to the small number of potential data points.
Since displaying all results is impracticable, the following Table~\ref{tab:feature} holds the best performing combination, among a sample of other combinations. 
\begin{table}[!htb]
    \centering
    \begin{tabular}{ll}
    \toprule
    \textbf{Feature Combination} & \textbf{F1-Score} \\ 
    \midrule
    QRT & 0.69 \\
    QT & 0.69 \\
    QW       & 0.72 \\
    \#T    & 0.68 \\
    QW, Loc                   & 0.72 \\
    QRT, QW                   & 0.72 \\
    QW, Loc                   & 0.73 \\
    QRT, QW, Loc              & 0.75 \\
    \#T, Loc, QW, QRT, Pers & 0.77 \\
    \textbf{\#T, Loc, QW, QRT}         & \textbf{0.78} \\
    \midrule
    \textbf{All features} & 0.76 \\
    \bottomrule
    \end{tabular}
    \caption{Different Feature Combinations. (Question Word QW, Number of Token \#T, Location Loc, Person Pers, Query Resource Type QRT)}
    \label{tab:feature}
\end{table}
0.78 is the globally highest F1-Score among all combinations.
This optimum is achieved by combining the number of features used, thus we chose to display the one with the least features required.  
These features are namely \texttt{Number of Token}, \texttt{Location}, \texttt{Question Word} and \texttt{Query Resource Type}.
Note, the performance decreases by 2 percent, using all features. 
Adding other features beyond the optimal group seems to introduce noise. 
However, this is highly dependent on the particular set of questions that is used.

\subsection{Metasystem Performance}

The overall goal was to develop a metasystem that is able to perform better than the underlying systems to benefit from the multitude of existing QA research and development activities.
As shown in Table~\ref{tab:qald6systems}, the six underlying systems perform with an F-measure of 0.15 to 0.68 on QALD-6. 
An optimal selection algorithm (which would always choose the best performing QA system) would achieve 0.89 F-measure.
%, improving the best system performance by 0.10 points F-measure or 30.3\%. 
Our best performing metasystem, trained on the 100 questions alone using the PSt classifier and only four features--namely \texttt{\#T}, \texttt{Loc}, \texttt{QW} and \texttt{QRT}-- is able to improve the best single system performance by 14.1\% and reaches an F-measure of 0.78. This result supports our assumptions about the diversity of existing QA solutions and shows how a good feature design allows characterization and the effective use of QA systems. The overall results however also show clear weaknesses of existing QA solutions. In particular, questions which require solution modifiers (e.g., 9, 88, 17, 28, 33, 36, 49) remain a difficult problem that need to be tackled. 

\section{Conclusion and Summary}

The QA metasystem we have presented is able to outperform each single QA system using a feature-selection approach combined with multi-label classification. 
We were able to show that an effective combination of systems, features and classifiers can improve overall performance.

However, our system is still more than 0.10 points F-measure away from an optimal system selection. 
This gap exists due to a lack of training data since we had only 100 training instances respectively questions available. 
Thus, we welcome other QA system developers to implement webservices to foster more active research and increase the comparability of systems. 
We have actively begun research to sophisticate the benchmarking of QA systems.\footnote{\url{http://gerbil-qa.aksw.org/gerbil/}}.
Furthermore, we will look deeper into the issues of overfitting classifiers and finding more influential features in the future.

\textbf{Acknowledgments}
This work has been supported by Eurostars projects DIESEL (E!9367) and QAMEL (E!9725) as well as the European Union's H2020 research and innovation action HOBBIT (GA 688227). We also thank Christina Unger for providing us with the underlying datasets.

\bibliographystyle{abbrv}
\bibliography{llncs}

\input{qald6systems}

\end{document}

%% file: qald6systems.tex
%\tiny
{\small
\begin{longtable}{lp{.6\textwidth}ccccccc|cc}
\toprule
\textbf{Id} & \textbf{Question}                                                    & \rot{\textbf{KWGAnswer}} &  \rot{\textbf{NbFramework}} &  \rot{\textbf{PersianQA}} &  \rot{\textbf{SemGraphQA}} &  \rot{\textbf{UIQA WME}} &  \rot{\textbf{UTQA English}} &  \rot{\textbf{Optimal}} &  \rot{\textbf{Metasystem}} \\
\midrule
0	 &Who was the doctoral supervisor of Albert Einstein?	 &0.0	 &1.0	 &0.0	 &0.0	 &0.0	 &1.0	 &1.0	 &1.0\\
1	 &Did Kaurismäki ever win the Grand Prix at Cannes?	 &1.0	 &0.0	 &0.0	 &0.0	 &0.0	 &1.0	 &1.0	 &1.0\\
2	 &Who wrote the song Hotel California?	 &0.0	 &0.0	 &0.0	 &0.0	 &0.0	 &0.0	 &0.0	 &0.0\\
3	 &Who was on the Apollo 11 mission?	 &1.0	 &0.0	 &0.0	 &0.0	 &0.0	 &0.0	 &1.0	 &0.0\\
4	 &Which electronics companies were founded in Beijing?	 &1.0	 &0.06	 &1.0	 &0.06	 &0.06	 &1.0	 &1.0	 &1.0\\
5	 &What is in a chocolate chip cookie?	 &1.0	 &1.0	 &0.0	 &0.0	 &0.0	 &1.0	 &1.0	 &1.0\\
6	 &What is the atmosphere of the Moon composed of?	 &1.0	 &0.0	 &0.0	 &0.0	 &0.0	 &1.0	 &1.0	 &1.0\\
7	 &How many movies did Park Chan-wook direct?	 &1.0	 &1.0	 &0.0	 &1.0	 &0.0	 &0.0	 &1.0	 &1.0\\
8	 &Who are the developers of DBpedia?	 &1.0	 &1.0	 &1.0	 &1.0	 &0.85	 &1.0	 &1.0	 &1.0\\
9	 &Which Indian company has the most employees?	 &0.0	 &0.0	 &0.0	 &0.0	 &0.0	 &0.0	 &0.0	 &0.0\\
\rowcolor[HTML]{FF9494} 
10	 &What is the name of the school where Obama's wife studied?	 &0.0	 &1.0	 &0.0	 &0.0	 &0.0	 &0.66	 &1.0	 &0.66\\
11	 &Where does Piccadilly start?	 &0.0	 &0.0	 &0.0	 &0.0	 &0.0	 &0.0	 &0.0	 &0.0\\
12	 &What is the capital of Cameroon?	 &1.0	 &1.0	 &1.0	 &1.0	 &1.0	 &1.0	 &1.0	 &1.0\\
13	 &When did the Boston Tea Party take place?	 &1.0	 &1.0	 &0.0	 &0.0	 &0.0	 &1.0	 &1.0	 &1.0\\
14	 &Who played Gus Fring in Breaking Bad?	 &0.0	 &1.0	 &0.0	 &0.0	 &0.0	 &1.0	 &1.0	 &1.0\\
15	 &Who wrote Harry Potter?	 &0.66	 &1.0	 &0.0	 &1.0	 &0.0	 &1.0	 &1.0	 &1.0\\
16	 &Which actors play in Big Bang Theory?	 &0.5	 &1.0	 &0.0	 &0.0	 &0.0	 &1.0	 &1.0	 &1.0\\
17	 &What is the largest country in the world?	 &0.0	 &0.0	 &0.0	 &0.0	 &0.0	 &1.0	 &1.0	 &1.0\\
18	 &Who is the most powerful Jedi?	 &0.0	 &1.0	 &1.0	 &1.0	 &0.0	 &1.0	 &1.0	 &1.0\\
19	 &How many goals did Pelé score?	 &1.0	 &1.0	 &1.0	 &1.0	 &0.0	 &0.0	 &1.0	 &1.0\\
\rowcolor[HTML]{FF9494} 
20	 &Who is the president of Eritrea?	 &1.0	 &1.0	 &0.0	 &0.03	 &1.0	 &0.66	 &1.0	 &0.66\\
21	 &Which computer scientist won an oscar?	 &0.0	 &0.0	 &0.0	 &0.0	 &0.0	 &1.0	 &1.0	 &1.0\\
22	 &Who created Family Guy?	 &1.0	 &1.0	 &0.0	 &1.0	 &0.0	 &1.0	 &1.0	 &1.0\\
\rowcolor[HTML]{FF9494} 
23	 &How many people live in Poland?	 &1.0	 &1.0	 &0.0	 &0.0	 &0.0	 &0.0	 &1.0	 &0.0\\
24	 &To which party does the mayor of Paris belong?	 &1.0	 &1.0	 &0.0	 &1.0	 &0.0	 &1.0	 &1.0	 &1.0\\
25	 &Who does the voice of Bart Simpson?	 &0.0	 &0.0	 &0.0	 &0.0	 &0.0	 &1.0	 &1.0	 &1.0\\
\rowcolor[HTML]{FF9494} 
26	 &Who composed the soundtrack for Cameron's Titanic?	 &1.0	 &0.0	 &0.0	 &0.0	 &0.0	 &0.0	 &1.0	 &0.0\\
27	 &When did Boris Becker end his active career?	 &0.0	 &1.0	 &0.0	 &0.0	 &0.0	 &0.0	 &1.0	 &1.0\\
28	 &Show me all basketball players that are higher than 2 meters.	 &0.0	 &0.0	 &0.0	 &0.0	 &0.0	 &1.0	 &1.0	 &1.0\\
29	 &What country is Sitecore from?	 &1.0	 &1.0	 &1.0	 &0.0	 &0.0	 &1.0	 &1.0	 &1.0\\
\rowcolor[HTML]{FF9494} 
30	 &Which country was Bill Gates born in?	 &1.0	 &0.0	 &0.0	 &0.0	 &1.0	 &0.0	 &1.0	 &0.0\\
31	 &Who developed Slack?	 &1.0	 &0.0	 &0.0	 &0.0	 &0.0	 &1.0	 &1.0	 &1.0\\
32	 &In which city did Nikos Kazantzakis die?	 &1.0	 &0.66	 &0.0	 &0.66	 &0.0	 &1.0	 &1.0	 &1.0\\
33	 &How many grand-children did Jacques Cousteau have?	 &0.0	 &0.0	 &0.0	 &0.0	 &0.0	 &0.0	 &0.0	 &0.0\\
34	 &Which films did Stanley Kubrick direct?	 &1.0	 &1.0	 &0.96	 &1.0	 &1.0	 &1.0	 &1.0	 &1.0\\
35	 &Does Neymar play for Real Madrid?	 &1.0	 &0.0	 &0.0	 &0.0	 &0.0	 &1.0	 &1.0	 &1.0\\
36	 &How many seats does the home stadium of FC Porto have?	 &1.0	 &0.0	 &0.0	 &0.0	 &0.0	 &1.0	 &1.0	 &1.0\\
37	 &Show me all books in Asimov's Foundation series.	 &0.95	 &0.0	 &0.21	 &0.0	 &0.0	 &1.0	 &1.0	 &1.0\\
\rowcolor[HTML]{FF9494} 
38	 &Which movies star both Liz Taylor and Richard Burton?	 &0.95	 &1.0	 &0.0	 &0.0	 &0.35	 &0.77	 &1.0	 &0.77\\
39	 &In which city are the headquarters of the United Nations?	 &0.0	 &0.0	 &0.0	 &0.0	 &0.0	 &0.0	 &0.0	 &0.0\\
\rowcolor[HTML]{FF9494} 
40	 &In which city was the president of Montenegro born?	 &1.0	 &1.0	 &0.0	 &0.0	 &1.0	 &0.66	 &1.0	 &0.66\\
\rowcolor[HTML]{FF9494} 
41	 &Which writers studied in Istanbul?	 &0.21	 &0.0	 &0.0	 &0.25	 &0.33	 &0.22	 &0.33	 &0.22\\
42	 &Who is the mayor of Paris?	 &1.0	 &1.0	 &1.0	 &1.0	 &1.0	 &1.0	 &1.0	 &1.0\\
43	 &What is the full name of Prince Charles?	 &1.0	 &0.0	 &0.0	 &1.0	 &0.0	 &1.0	 &1.0	 &1.0\\
\rowcolor[HTML]{FF9494} 
44	 &What is the longest river in China?	 &0.0	 &1.0	 &0.0	 &0.0	 &0.02	 &0.0	 &1.0	 &0.0\\
45	 &Who discovered Ceres?	 &1.0	 &1.0	 &0.0	 &1.0	 &0.0	 &0.0	 &1.0	 &1.0\\
46	 &When did princess Diana die?	 &1.0	 &1.0	 &0.0	 &0.0	 &1.0	 &1.0	 &1.0	 &1.0\\
47	 &What do ants eat?	 &1.0	 &1.0	 &1.0	 &1.0	 &0.0	 &1.0	 &1.0	 &1.0\\
\rowcolor[HTML]{FF9494} 
48	 &Who is the host of the BBC Wildlife Specials?	 &1.0	 &1.0	 &0.0	 &1.0	 &0.0	 &0.0	 &1.0	 &0.0\\
49	 &How many moons does Mars have?	 &0.0	 &0.0	 &0.0	 &0.0	 &0.0	 &0.0	 &0.0	 &0.0\\
50	 &What was the first Queen album?	 &0.02	 &0.0	 &0.0	 &0.0	 &0.0	 &0.0	 &0.02	 &0.02\\
51	 &Did Elvis Presley have children?	 &1.0	 &0.0	 &0.0	 &0.0	 &0.0	 &0.0	 &1.0	 &1.0\\
52	 &Give me a list of all Canadians that reside in the U.S.	 &0.0	 &0.0	 &0.0	 &0.0	 &0.0	 &1.0	 &1.0	 &1.0\\
53	 &Where is Syngman Rhee buried?	 &1.0	 &1.0	 &0.0	 &0.0	 &1.0	 &1.0	 &1.0	 &1.0\\
54	 &In which countries do people speak Japanese?	 &1.0	 &0.5	 &0.0	 &0.0	 &0.0	 &1.0	 &1.0	 &1.0\\
55	 &Who is the king of the Netherlands?	 &0.0	 &0.66	 &0.0	 &0.0	 &0.0	 &1.0	 &1.0	 &1.0\\
56	 &When did the Dodo become extinct?	 &1.0	 &1.0	 &0.0	 &0.0	 &0.0	 &1.0	 &1.0	 &1.0\\
57	 &Show me all Czech movies.	 &0.89	 &1.0	 &0.0	 &0.0	 &0.0	 &1.0	 &1.0	 &1.0\\
\rowcolor[HTML]{FF9494} 
58	 &Which rivers flow into the North Sea?	 &1.0	 &0.43	 &0.45	 &0.45	 &0.45	 &0.22	 &1.0	 &0.22\\
\rowcolor[HTML]{FF9494} 
59	 &When did Operation Overlord commence?	 &1.0	 &0.66	 &0.0	 &1.0	 &0.0	 &1.0	 &1.0	 &0.66\\
60	 &Where do the Red Sox play?	 &1.0	 &0.0	 &0.0	 &0.0	 &0.0	 &1.0	 &1.0	 &1.0\\
61	 &In which time zone is Rome?	 &1.0	 &1.0	 &1.0	 &1.0	 &0.0	 &1.0	 &1.0	 &1.0\\
62	 &Give me a list of all critically endangered birds.	 &0.0	 &0.0	 &0.0	 &0.0	 &0.0	 &1.0	 &1.0	 &1.0\\
\rowcolor[HTML]{FF9494} 
63	 &How much did the Lego Movie cost?	 &0.5	 &0.0	 &0.0	 &0.0	 &0.0	 &1.0	 &1.0	 &0.5\\
64	 &What was the original occupation of the inventor of Lego?	 &0.0	 &0.0	 &0.0	 &0.66	 &1.0	 &1.0	 &1.0	 &1.0\\
65	 &Which countries have more than ten volcanoes?	 &0.0	 &0.87	 &0.0	 &0.0	 &0.0	 &1.0	 &1.0	 &1.0\\
66	 &Show me all U.S. states.	 &0.0	 &1.0	 &0.0	 &0.0	 &0.0	 &1.0	 &1.0	 &1.0\\
67	 &Who wrote the Game of Thrones theme?	 &0.0	 &0.0	 &0.0	 &0.0	 &0.0	 &1.0	 &1.0	 &1.0\\
68	 &How many calories does a baguette have?	 &0.0	 &1.0	 &0.0	 &0.0	 &0.0	 &0.0	 &1.0	 &1.0\\
69	 &Can you cry underwater?	 &1.0	 &1.0	 &1.0	 &1.0	 &0.0	 &1.0	 &1.0	 &1.0\\
\rowcolor[HTML]{FF9494} 
70	 &Which companies produce hovercrafts?	 &1.0	 &1.0	 &1.0	 &1.0	 &0.0	 &0.0	 &1.0	 &0.0\\
71	 &How many emperors did China have?	 &0.0	 &0.0	 &0.0	 &0.0	 &0.0	 &1.0	 &1.0	 &1.0\\
72	 &Show me hiking trails in the Grand Canyon where there's no danger of flash floods.	 &0.0	 &0.0	 &0.0	 &0.0	 &0.0	 &1.0	 &1.0	 &1.0\\
73	 &In which ancient empire could you pay with cocoa beans?	 &1.0	 &1.0	 &0.0	 &0.0	 &0.0	 &1.0	 &1.0	 &1.0\\
74	 &How did Michael Jackson die?	 &0.0	 &0.0	 &0.0	 &0.0	 &1.0	 &0.0	 &1.0	 &1.0\\
75	 &Which space probes were sent into orbit around the sun?	 &0.0	 &0.0	 &0.0	 &0.0	 &0.0	 &0.72	 &0.72	 &0.72\\
76	 &When was Coca Cola invented?	 &1.0	 &1.0	 &1.0	 &1.0	 &0.0	 &0.0	 &1.0	 &1.0\\
77	 &What is the biggest stadium in Spain?	 &0.01	 &0.0	 &0.0	 &0.0	 &0.01	 &1.0	 &1.0	 &1.0\\
78	 &On which day is Columbus Day?	 &0.0	 &0.0	 &0.0	 &0.0	 &0.0	 &1.0	 &1.0	 &1.0\\
79	 &How short is the shortest active NBA player?	 &0.0	 &0.0	 &0.0	 &0.0	 &0.0	 &0.0	 &0.0	 &0.0\\
\rowcolor[HTML]{FF9494} 
80	 &Whom did Lance Bass marry?	 &1.0	 &1.0	 &1.0	 &0.0	 &1.0	 &0.66	 &1.0	 &0.66\\
81	 &What form of government does Russia have?	 &0.0	 &1.0	 &0.0	 &0.0	 &0.0	 &1.0	 &1.0	 &1.0\\
82	 &What movies does Jesse Eisenberg play in?	 &1.0	 &0.98	 &1.0	 &0.0	 &0.98	 &0.0	 &1.0	 &1.0\\
83	 &What color expresses loyalty?	 &0.0	 &0.0	 &0.0	 &0.0	 &0.0	 &1.0	 &1.0	 &1.0\\
84	 &Show me all museums in London.	 &1.0	 &1.0	 &0.0	 &0.0	 &0.0	 &1.0	 &1.0	 &1.0\\
85	 &Give me all South American countries.	 &1.0	 &1.0	 &0.0	 &0.0	 &0.0	 &1.0	 &1.0	 &1.0\\
86	 &Who is the coach of Ankara's ice hockey team?	 &0.0	 &0.0	 &0.0	 &0.0	 &0.0	 &1.0	 &1.0	 &1.0\\
87	 &Who is the son of Sonny and Cher?	 &0.0	 &0.0	 &0.0	 &0.0	 &1.0	 &1.0	 &1.0	 &1.0\\
88	 &What are the five boroughs of New York?	 &0.08	 &0.0	 &0.0	 &0.0	 &0.0	 &0.28	 &0.28	 &0.28\\
89	 &Show me Hemingway's autobiography.	 &0.0	 &0.0	 &0.0	 &0.0	 &0.0	 &1.0	 &1.0	 &1.0\\
90	 &What kind of music did Lou Reed play?	 &0.0	 &0.0	 &0.0	 &0.0	 &0.0	 &1.0	 &1.0	 &1.0\\
91	 &In which city does Sylvester Stallone live?	 &1.0	 &1.0	 &0.66	 &0.0	 &0.0	 &1.0	 &1.0	 &1.0\\
92	 &Who was Vincent van Gogh inspired by?	 &1.0	 &0.0	 &1.0	 &0.0	 &0.0	 &1.0	 &1.0	 &1.0\\
93	 &What are the names of the Teenage Mutant Ninja Turtles?	 &0.0	 &0.0	 &0.0	 &0.0	 &0.0	 &1.0	 &1.0	 &1.0\\
94	 &What are the zodiac signs?	 &0.0	 &0.0	 &0.0	 &0.0	 &0.0	 &1.0	 &1.0	 &1.0\\
\rowcolor[HTML]{FF9494} 
95	 &What languages do they speak in Pakistan?	 &1.0	 &0.37	 &0.37	 &0.0	 &0.0	 &0.32	 &1.0	 &0.32\\
96	 &Who became president after JFK died?	 &0.0	 &0.0	 &0.0	 &0.0	 &0.0	 &0.0	 &0.0	 &0.0\\
97	 &In what city is the Heineken brewery?	 &1.0	 &0.0	 &0.0	 &1.0	 &0.0	 &0.0	 &1.0	 &1.0\\
98	 &What is Elon Musk famous for?	 &1.0	 &1.0	 &0.0	 &0.0	 &0.0	 &0.44	 &1.0	 &1.0\\
99	 &What is Batman's real name?	 &0.0	 &0.0	 &0.0	 &0.0	 &0.0	 &1.0	 &1.0	 &1.0\\
\midrule
& \textbf{Average F-measure }  &0.54 &0.48 &0.17 &0.22 &0.15 &0.68 &0.89 & \textbf{0.78}   \\\bottomrule
\caption{Performance of QA systems on QALD-6.} 
\label{tab:qald6systems}
\end{longtable}
}

%% file: llncs.bbl
\begin{thebibliography}{10}

\bibitem{YodaQA}
P.~Baudis and J.~Sediv{\'{y}}.
\newblock Modeling of the question answering task in the yodaqa system.
\newblock In {\em Experimental {IR} Meets Multilinguality, Multimodality, and
  Interaction - 6th International Conference of the {CLEF} Association, {CLEF}
  2015, Toulouse, France, September 8-11, 2015, Proceedings}, pages 222--228,
  2015.

\bibitem{beaumont2015semgraphqa}
R.~Beaumont, B.~Grau, and A.-L. Ligozat.
\newblock Semgraphqa at qald-5: Limsi participation at qald-5 at clef.
\newblock In {\em CLEF (Working Notes)}, 2015.

\bibitem{both2014service}
A.~Both, A.-C.~N. Ngonga, R.~Usbeck, D.~Lukovnikov, C.~Lemke, and M.~Speicher.
\newblock A service-oriented search framework for full text, geospatial and
  semantic search.
\newblock In {\em SEMANTiCS}, 2014.

\bibitem{cheng2010bayes}
W.~Cheng, E.~H{\"u}llermeier, and K.~J. Dembczynski.
\newblock Bayes optimal multilabel classification via probabilistic classifier
  chains.
\newblock In {\em Proceedings of the 27th international conference on machine
  learning (ICML-10)}, pages 279--286, 2010.

\bibitem{qallme}
O.~Ferrandez, C.~Spurk, M.~Kouylekov, I.~Dornescu, S.~Ferrandez, M.~Negri,
  R.~Izquierdo, D.~Tomas, C.~Orasan, G.~Neumann, et~al.
\newblock The qall-me framework: A specifiable-domain multilingual question
  answering architecture.
\newblock {\em Web semantics: Science, services and agents on the world wide
  web}, 9(2):137--145, 2011.

\bibitem{treo}
A.~Freitas, J.~G. Oliveira, E.~Curry, S.~O’Riain, and J.~C.~P. da~Silva.
\newblock Treo: combining entity-search, spreading activation and semantic
  relatedness for querying linked data.
\newblock In {\em 1st Workshop on Question Answering over Linked Data
  (QALD-1)}, 2011.

\bibitem{guo2011multi}
Y.~Guo and S.~Gu.
\newblock Multi-label classification using conditional dependency networks.
\newblock In {\em IJCAI Proceedings-International Joint Conference on
  Artificial Intelligence}, volume~22, page 1300, 2011.

\bibitem{qasurvey}
K.~Höffner, S.~Walter, E.~Marx, J.~Lehmann, A.~Ngonga, and R.~Usbeck.
\newblock Overcoming challenges of semantic question answering in the semantic
  web.
\newblock {\em Semantic Web Journal}, 2016.

\bibitem{Kolomiyets:2011}
O.~Kolomiyets and M.-F. Moens.
\newblock A survey on question answering technology from an information
  retrieval perspective.
\newblock {\em Inf. Sci.}, 181(24):5412--5434, Dec. 2011.

\bibitem{jl_2014/swj_dbpedia}
J.~Lehmann, R.~Isele, M.~Jakob, A.~Jentzsch, D.~Kontokostas, P.~Mendes,
  S.~Hellmann, M.~Morsey, P.~van Kleef, S.~Auer, and C.~Bizer.
\newblock {DB}pedia - a large-scale, multilingual knowledge base extracted from
  wikipedia.
\newblock {\em Semantic Web Journal}, 2014.

\bibitem{PowerAqua}
V.~Lopez, M.~Fern{\'{a}}ndez, E.~Motta, and N.~Stieler.
\newblock {PowerAqua: Supporting users in querying and exploring the Semantic
  Web}.
\newblock {\em Semantic Web Journal}, 3:249--265, 2012.

\bibitem{DBLP:journals/semweb/LopezUSM11}
V.~Lopez, V.~S. Uren, M.~Sabou, and E.~Motta.
\newblock Is question answering fit for the semantic web?: {A} survey.
\newblock {\em Semantic Web Journal}, 2(2):125--155, 2011.

\bibitem{stanfordcorenlp}
C.~D. Manning, M.~Surdeanu, J.~Bauer, J.~Finkel, S.~J. Bethard, and
  D.~McClosky.
\newblock The {Stanford} {CoreNLP} natural language processing toolkit.
\newblock In {\em 52nd ACL: System Demonstrations}, pages 55--60, 2014.

\bibitem{openQA}
E.~Marx, R.~Usbeck, A.-C.~N. Ngonga, K.~H{\"o}ffner, J.~Lehmann, and S.~Auer.
\newblock Towards an open question answering architecture.
\newblock In {\em SEMANTiCS}, 2014.

\bibitem{mazzeo2016canali}
G.~M. Mazzeo and C.~Zaniolo.
\newblock Canali: A system for answering controlled natural language questions
  on rdf knowledge bases, 2016.

\bibitem{qanus}
J.-P. Ng and M.-Y. Kan.
\newblock Qanus: An open-source question-answering platform.
\newblock {\em arXiv preprint arXiv:1501.00311}, 2015.

\bibitem{read2008pruned}
J.~Read.
\newblock A pruned problem transformation method for multi-label
  classification.
\newblock In {\em Proc. 2008 New Zealand Computer Science Research Student
  Conference (NZCSRS 2008)}, volume 143150, 2008.

\bibitem{read2014efficient}
J.~Read, L.~Martino, and D.~Luengo.
\newblock Efficient monte carlo methods for multi-dimensional learning with
  classifier chains.
\newblock {\em Pattern Recognition}, 47(3):1535--1546, 2014.

\bibitem{read2011classifier}
J.~Read, B.~Pfahringer, G.~Holmes, and E.~Frank.
\newblock Classifier chains for multi-label classification.
\newblock {\em Machine learning}, 85(3):333--359, 2011.

\bibitem{MEKA}
J.~Read, P.~Reutemann, B.~Pfahringer, and G.~Holmes.
\newblock {MEKA}: A multi-label/multi-target extension to {Weka}.
\newblock {\em Journal of Machine Learning Research}, 17(21):1--5, 2016.

\bibitem{Schapire:2000:BBS:347709.347732}
R.~E. Schapire and Y.~Singer.
\newblock Boostexter: A boosting-based systemfor text categorization.
\newblock {\em Mach. Learn.}, 39(2-3):135--168, May 2000.

\bibitem{SINA_WebSemantic}
S.~Shekarpour, E.~Marx, A.-C.~N. Ngomo, and S.~Auer.
\newblock Sina: Semantic interpretation of user queries for question answering
  on interlinked data.
\newblock {\em Journal of Web Semantics}, 2014.

\bibitem{singhqanary}
K.~Singh, A.~Both, D.~Diefenbach, S.~Shekarpour, D.~Cherix, and C.~Lange15.
\newblock Qanary--the fast track to creating a question answering system with
  linked data technology.
\newblock In {\em ESWC}, 2016.

\bibitem{tsoumakas2011random}
G.~Tsoumakas, I.~Katakis, and I.~Vlahavas.
\newblock Random k-labelsets for multilabel classification.
\newblock {\em IEEE Transactions on Knowledge and Data Engineering},
  23(7):1079--1089, 2011.

\bibitem{tbsl}
C.~Unger, L.~B{\"{u}}hmann, J.~Lehmann, A.~N. Ngomo, D.~Gerber, and P.~Cimiano.
\newblock {Template-based question answering over {RDF} data}.
\newblock In {\em 21st WWW conference}, pages 639--648, 2012.

\bibitem{qald4}
C.~Unger, C.~Forascu, V.~Lopez, A.~N. Ngomo, E.~Cabrio, P.~Cimiano, and
  S.~Walter.
\newblock Question answering over linked data {(QALD-4)}.
\newblock In {\em {CLEF}}, pages 1172--1180, 2014.

\bibitem{qald5}
C.~Unger, C.~Forascu, V.~Lopez, A.~N. Ngomo, E.~Cabrio, P.~Cimiano, and
  S.~Walter.
\newblock Question answering over linked data {(QALD-5)}.
\newblock In {\em Working Notes of {CLEF} 2015 - Conference and Labs of the
  Evaluation forum, Toulouse, France, September 8-11, 2015.}, 2015.

\bibitem{qald6}
C.~Unger, A.~Ngonga, and E.~Cabrio.
\newblock 6th open challenge on question answering over linked data (qald-6).
\newblock In {\em The Semantic Web: ESWC 2016 Challenges.}, 2016.

\bibitem{HAWK_NLIWOD_2015}
R.~Usbeck, E.~K{\"o}rner, and A.-C. {Ngonga Ngomo}.
\newblock Answering boolean hybrid questions with hawk.
\newblock In {\em NLIWOD workshop at International Semantic Web Conference
  (ISWC), including erratum and changes}, 2015.

\bibitem{hawk}
R.~Usbeck, A.-C. Ngomo, L.~Bühmann, and C.~Unger.
\newblock Hawk – hybrid question answering using linked data.
\newblock In {\em The Semantic Web. Latest Advances and New Domains}, volume
  9088 of {\em Lecture Notes in Computer Science}, pages 353--368. Springer
  International Publishing, 2015.

\bibitem{utqa}
A.~P.~B. Veyseh.
\newblock Cross-lingual question answering using common semantic space.
\newblock In {\em Proceedings of TextGraphs@NAACL-HLT 2016}, pages 15--19,
  2016.

\bibitem{Xser}
K.~Xu, Y.~Feng, S.~Huang, and D.~Zhao.
\newblock Question answering via phrasal semantic parsing.
\newblock In {\em Experimental {IR} Meets Multilinguality, Multimodality, and
  Interaction - 6th International Conference of the {CLEF} Association, {CLEF}
  2015, Toulouse, France, September 8-11, 2015, Proceedings}, pages 414--426,
  2015.

\bibitem{oaqa}
Z.~Yang, E.~Garduno, Y.~Fang, A.~Maiberg, C.~McCormack, and E.~Nyberg.
\newblock Building optimal information systems automatically: Configuration
  space exploration for biomedical information systems.
\newblock In {\em 22nd ACM CIKM}, pages 1421--1430. ACM, 2013.

\bibitem{zhang2007ml}
M.-L. Zhang and Z.-H. Zhou.
\newblock Ml-knn: A lazy learning approach to multi-label learning.
\newblock {\em Pattern recognition}, 40(7):2038--2048, 2007.

\end{thebibliography}
